\documentclass[10pt,conference]{IEEEtran}

\usepackage[utf8]{inputenc}
\usepackage[T1]{fontenc}    
\usepackage{hyperref}       
\usepackage{url}            
\usepackage{booktabs}       
\usepackage{amsfonts}       
\usepackage{nicefrac}       
\usepackage{microtype}      
\usepackage{lipsum}		
\usepackage{graphicx}
\usepackage[natbib=true, sorting=none]{biblatex}
\usepackage{doi}
\usepackage{comment}

\usepackage{xcolor}
\usepackage{physics}
\usepackage{makecell}
\usepackage{multirow}
\usepackage{graphicx}
\usepackage{subcaption}

\usepackage{physics}
\usepackage{algorithm}
\usepackage{algpseudocode}
\usepackage[capitalise]{cleveref}
\begin{document}

\title{Building Continuous Quantum-Classical Bayesian Neural Networks for a Classical Clinical Dataset}
\author{ Alona Sakhnenko\IEEEauthorrefmark{1}, Julian Sikora\IEEEauthorrefmark{1}\IEEEauthorrefmark{3},Jeanette Miriam Lorenz\IEEEauthorrefmark{1}\IEEEauthorrefmark{2} \\
	\IEEEauthorblockA{\IEEEauthorrefmark{1}Fraunhofer Institute for Cognitive Systems IKS,  Munich, Germany}
	\IEEEauthorblockA{\IEEEauthorrefmark{2}Ludwig-Maximilian University, Munich, Germany}
        \IEEEauthorblockA{\IEEEauthorrefmark{3}Technical University of Munich, Munich, Germany}
	\texttt{\{alona.sakhnenko, julian.sikora, jeanette.miriam.lorenz\}@iks.fraunhofer.de} \\
}
\date{\today}

\IEEEoverridecommandlockouts
\IEEEpubid{\makebox[\columnwidth]{979-8-4007-1798-7/24/06~\copyright2024 ReAQCT \hfill}
\hspace{\columnsep}\makebox[\columnwidth]{ }}

\maketitle
\begin{abstract}
In this work, we are introducing a Quantum-Classical Bayesian Neural Network (QCBNN) that is capable to perform uncertainty-aware classification of classical medical dataset. This model is a symbiosis of a classical Convolutional NN that performs ultra-sound image processing and a quantum circuit that generates its stochastic weights, within a Bayesian learning framework. To test the utility of this idea for the possible future deployment in the medical sector we track multiple behavioral metrics that capture both predictive performance as well as model's uncertainty. It is our ambition to create a hybrid model that is capable to classify samples in a more uncertainty aware fashion, which will advance the trustworthiness of these models and thus bring us step closer to utilizing them in the industry. We test multiple setups for quantum circuit for this task, and our best architectures display bigger uncertainty gap between correctly and incorrectly identified samples than its classical benchmark at an expense of a slight drop in predictive performance. The innovation of this paper is two-fold: (1) combining of different approaches that allow the stochastic weights from the quantum circuit to be continues thus allowing the model to classify application-driven dataset; (2) studying architectural features of quantum circuit that make-or-break these models, which pave the way into further investigation of more informed architectural designs.
\end{abstract}
\begin{IEEEkeywords}
Quantum Bayesian Neural Networks, QCBNN, Hybrid models, PQC architecture study
\end{IEEEkeywords}

\section{Introduction}
Neural Networks (NNs) have achieved incredible results in many fields of Machine Learning. However, some practical applications, such as medical AI, have strict reliability requirements where NNs may lead to catastrophic failure if they cannot reliably estimate the confidence interval of their output. Bayesian learning provides a well-established framework to train and analyse uncertainty-aware models, and Bayesian NNs (BNNs) are stochastic NNs that yield the benefits of Bayesian learning in a deep learning setup. The success of these models depend on a good choice of both a NN architecture and stochastic distribution family.

Quantum Computing (QC) is an emerging technology that has been rapidly gaining momentum in the last years. Many research initiatives have been directed towards investigating early applications of this technology in a real-world scenario. There is, however, one substantial challenge for application-driven approaches, which is Noisy Intermediate-Scale Quantum (NISQ) devices that are available today. Unlike fault-tolerant computers that promise provable advantages, such as with Shor and HHL algorithms, the advantages of NISQ algorithms are still being explored.

Early proof-of-concept applications of NISQ for Machine Learning (ML) problems have shown a lot of potential of boosting capacities of ML models. These applications have provided both \textit{theory-driven} insights, e.g. Quantum Neural Networks showed higher effective dimensions on an investigated problem~\cite{Abbas_2021}, Quantum Kernels allow for better generalization on specific datasets~\cite{Huang_2021}, Born Machines are capable of representing distributions that are hard to sample classically~\cite{Coyle_2020}, as well as \textit{empirically-driven} results, e.g. boosted performance of a hybrid Quantum-Classical Generative NN for high-resolution images~\cite{rudolph2022generation}, boosted performance of a hybrid Quantum-Classical Convolutional NN (QCCNN) on medical dataset while requiring less training parameters than the classical benchmark~\cite{matic2022quantumclassical}.

In this work, we explore how we can leverage NISQ devices to create uncertainty-aware models by utilizing quantum circuits for Bayesian learning and how this setup compares to its classical analog. We concentrate on application-driven approach that is suitable for potential future deployment in medical AI sector. As a proof-of-concept use-case we select a publicly available Breast Ultrasound Images (BreastMNIST) dataset. We combine several architectural ideas together, e.g. \textit{binary} quantum-classical BNN~\cite{binary_qBNN}, \textit{continuous} quantum Generative Adversarial Networks~\cite{cont_distribution} that propose an idea to sample beyond the discrete distribution as well as QCCNN~\cite{matic2022quantumclassical} that proved useful in a hybrid quantum-classical setting on the same dataset. We create a Quantum-Classical (Convolutional) BNN (QCBNN) that processes ultrasound images classically, while the stochastic continuous weights are generated by a quantum circuit. We track multiple behavioral metrics of the model that capture both its predictive performance and its uncertainty. The proposed model shows a bigger gap in confidence of correctly and incorrectly identified samples than its classical benchmark, which means our model showed higher uncertainty on datapoints that were misclassified and hence increasing the trustworthiness of these models. 
Furthermore, we perform a systematic study of the effects of different parametrized quantum circuit (PQC) architectures from the literature, and build tailor-made PQCs for this task that increase the confidence gap. We provide the analysis of these architectures from a standpoint of different metrics, including confidence errors and ensemble sizes of correctly and incorrectly identified samples, model's calibration and the shape of weights distribution, which offers insights on how different architectural elements influence models' behaviour.

\section{Related work}\label{sec:related}

One of the early explorations into utility of fault-tolerant QC for Bayesian learning has been performed by \citet{Zhao_2019}. This work proposed to use HHL algorithm to accelerate matrix inversion procedure in the context of Gaussian Processes (GP). GPs have many beneficial properties, but, due to scaling bottlenecks, these models are not often used in practice. HHL algorithm can be leveraged to speed up the inference procedure of these models and, hence, improve their applicability. \citet{berner2021quantum} proposed to utilize another fault-tolerant algorithms for inner product estimation to accelerate training and inference of BNNs. Both of these algorithms have shown their usability on small examples, however, making them suitable for real-world instances would require either viable NISQ-adaptations of these algorithms or fault-tolerant~QCs. 

Other works looked into Bayesian learning in NISQ era. Work by \citet{9759399} showed the utility of Bayesian learning for training Quantum NNs. The authors showed that their method yield a better generalization (based on the study of model's capacity), and provided additional information on epistemic uncertainty. In our work, we investigate Bayesian learning for stochastic models in contrast to deterministic models \cite{9759399}, which is a more common method for building uncertainty-aware models. One of the first attempts to translate stochastic models into the quantum realm was done by \citet{binary_qBNN}, where the authors introduced a quantum binary BNN that utilizes a Born machine to generate binary stochastic weights. The authors show that this approach allows for an easier way to optimize over discrete distributions, and the utility of it was shown on a simple regression example. Scaling this idea to more complex problems might prove difficult because of limited capacity of binary NNs. Hence, in this work we investigate how we can extend this approach to sample from continuous distributions and its potential benefits on a medical use-case.  

\citet{matic2022quantumclassical} demonstrated a benefit of using quantum-classical setup to boost performance on the same BreastMNIST dataset. In their work, the authors propose a Quantum-Classical Convolutional NNs, in which PQC performed a convolution operation and the expectation values of each qubit were propagated further into a fully-connected layer. This model achieved a better performance than its classical benchmark while requiring less trainable parameters. Subsequent work~\cite{monnet2023pooling} tested additionally different quantum pooling techniques, which allowed to boost performance even further. In our work, we build upon the architecture that was used in \cite{matic2022quantumclassical}, but change the role of a PQC to sample from weights distribution instead of performing convolution operation. We concentrate on the most successful PQC architectures to test their utility for our task. In our work, we extend the family of investigated PQCs and investigate which of their features are beneficial in our setup. 

\section{Background}
In this section, we provide a detailed description of important notions from classical (\cref{sec:BNN}) and quantum machine learning (\cref{sec:VI_on_QC}-\ref{sec:binary_BNN}) that are essential building blocks of QCBNN.

\subsection{Bayesian Neural Networks}\label{sec:BNN}
Bayesian Neural Networks (BNNs) are stochastic NN trained using Bayesian inference, which comprise of (1) a stochastic model of a chosen parametrization prior $p(\mathbf{w})$ and a prior of confidence $p(y|x,\mathbf{w})$, which result in a following posterior on training data $D$: 
\begin{equation}~\label{eq:posterior}
    p(\mathbf{w}|D) = \frac{p(D_y|D_x, \mathbf{w})p(\mathbf{w})}{\int_{\mathbf{w}} p(D_y|D_x, \mathbf{w}')p(\mathbf{w}')d \mathbf{w}'},
\end{equation}
and (2) a functional model $\Phi_{\mathbf{w}}(x)$, which in this case is a NN. Prediction process with BNNs can be represented in a general case as \cite{bnn_overview}:
\begin{align} \label{eq:bnn_general}
    &\mathbf{\mathbf{w}} \sim p(\mathbf{\mathbf{w}}|D) \nonumber \\
    &\mathbf{y} = \Phi_{\mathbf{\mathbf{w}}}(\mathbf{x}) + \epsilon,
\end{align}
where $\mathbf{x}$ is an input, $\mathbf{y}$ is a corresponding label and $\epsilon$ is an approximation error between output for a functional model output and the label. To perform classification with BNNs, we can compute an average model prediction over the ensemble of $N$ different outputs. To begin with, we compute the probability of each class as follows:
\begin{equation}
    \hat{\mathbf{p}} = \frac{1}{N} \sum_{i=0}^N \Phi_{\mathbf{\mathbf{w}}_i}(\mathbf{x}),
\end{equation}
from which we can compute the final prediction as $\hat{\mathbf{y}} = \arg \max_i p_i \in \hat{\mathbf{p}}$.

The challenge in working with BNNs resides in the fact that sampling directly from the posterior distribution (Equation~\ref{eq:posterior}) is difficult due to intractability of the evidence $\int_{\mathbf{w}} p(D_y|D_x, \mathbf{w}')p(\mathbf{w}')d \mathbf{w}'$ \cite{bnn_overview}. Instead, most commonly, we can either build a Markov chain with a stationary distribution being the posterior (Markov chain Monte Carlo methods) or select a tractable distribution and adjust its parameters to approximate the posterior (Variational inference). Due to its scalability, Variational inference (VI) has gain popularity within the ML community and is therefore the focus of this work. For VI we select a family of tractable distributions $q_{\theta}(\mathbf{x}) \in \mathcal{Q}$ and tweak the variational parameters $\theta$ to minimize the KL divergence between two distributions:

\begin{equation}\label{eq:KL_divergence}
    \mathbb{KL} [q_{\theta}(\mathbf{w}|D)|| p(\mathbf{w}|D)] 
    = \mathbb{E}_{\mathbf{w} \sim q_{\theta}(\mathbf{w}|D)} \Big[ \log \frac{q_{\theta}(\mathbf{w}|D)}{p(\mathbf{w}|D)} \Big] 
\end{equation}

\subsection{Variational Inference and Adversarial Training on a QC}\label{sec:VI_on_QC}

Born machines offer a way to model a distribution using a quantum state $\ket{\psi(\theta,x)}$, that encodes an input vector $x$ and is parameterized by $\theta$. These models generate bitstrings with probabilities $q_{\theta}(w|x) = |\braket{w}{\psi(\theta,x)}|^2$~\cite{Benedetti_2021}. The power of this method to model intractable distributions has been shown in \cite{coyle2020quantum}, and its generalization capabilities have be investigated in \cite{born_generalization}.

A VI training loop for Born machines was proposed by \citet{Benedetti_2021}. As mentioned above, KL divergence is a popular objective function for VI, which we can simplify as follows:
\begin{align}
    \mathbb{KL} &[q_{\theta}(\mathbf{w}|D)|| p(\mathbf{w}|D)] \\
    & = \mathbb{E}_{\mathbf{w} \sim q_{\theta}(\mathbf{w}|D)} \Big[ \log \frac{q_{\theta}(\mathbf{w}|D)}{p(\mathbf{w}|D)} \Big]  \nonumber \\
    & = \mathbb{E}_{\mathbf{w} \sim q_{\theta}(\mathbf{w}|D)} \Big[ \log \frac{q_{\theta}(\mathbf{w}|D) p(D)}{p(D|\mathbf{w}) p(\mathbf{w})} \Big] \nonumber \\
    & = \mathbb{E}_{\mathbf{w} \sim q_{\theta}(\mathbf{w}|D)} \Big[ \log \frac{q_{\theta}(\mathbf{w}|D) }{p(\mathbf{w})}  - \log p(D|\mathbf{w}) \Big] + const.
\end{align}
Unfortunately, we cannot estimate $\frac{q_{\theta}(\mathbf{w}|D)}{p(\mathbf{w})}$ directly on a quantum computer. To tackle this issue, adversarial learning can be utilized. We can construct a binary classifier $d_{\phi}(\mathbf{w}, D)$, which discriminates between samples from the true distribution and Born machine generated ones. The classifier objective function is defined as a cross-entropy:
\begin{equation}
    \mathcal{G}_{KL}(\phi; \theta) = \mathbb{E}_{\mathbf{w} \sim q_{\theta}(\mathbf{w}|D)}[\log d_{\phi}] + \mathbb{E}_{\mathbf{w} \sim q_{\theta}(\mathbf{w})}[\log( 1 -  d_{\phi})]
\end{equation}
In order to extract the odds of a data point coming from one of the two distributions, the classifier's output can be transformed using the logit function:
\begin{equation}
    \text{logit}(d_{\phi}(\mathbf{w}, D)) = \log \frac{d_{\phi}(\mathbf{w}, D)}{1 - d_{\phi}(\mathbf{w}, D)} \approx \log \frac{q_{\theta}(\mathbf{w}, D)}{p(\mathbf{w})}.
\end{equation}
Therefore, the final objective of the Born training is:
\begin{equation}\label{eq:Born_objective}
    \mathcal{L}_{KL}(\theta; \phi) =\mathbb{E}_{\mathbf{w} \sim q_{\theta}(\mathbf{w}|D)}[\text{logit}(d_{\phi}(\mathbf{w}, D)) - \log p (D|\mathbf{w})].
\end{equation}
During the overall training procedure, both a classifier and a Born machine are trained simultaneously as follows:
\begin{align}
    &\max_{\phi} \mathcal{G}_{KL} (\phi; \theta) \nonumber \\
    &\min_{\theta} \mathcal{L}_{KL}(\theta; \phi).
\end{align}

This adversarial training strategy is akin to the one commonly used for Generative Adversarial Networks (GANs). Quantum adaptation of these networks have been successfully applied to a variety of different use-cases \cite{rudolph2022generation, Zoufal_2019, jain2022hybrid}. \citet{cont_distribution} presented an approach that allows a quantum GANs to sample from a continuous distribution. Their approach suggests to inject classical noise into a PQC and a set of observables, whose expectation values serve as an output of the model. This method can be used to create continuous distributions for quantum VI method described above.

\subsection{Quantum-Classical Bayesian Neural Networks}\label{sec:binary_BNN}
As mentioned in \cref{sec:related}, \citet{binary_qBNN} proposed an initial architecture for QCBNNs. Their method followed the same techniques described in \cref{sec:VI_on_QC} and generated a binary distribution that proved its viability on a toy regression example in meta learning setting. The loss function of the model comprise of likelihood and adversarial terms as follows: 
\begin{equation}
    \mathcal{L} = -\alpha \mathbb{E}[\log p(D|\theta)] + \beta \mathcal{L}_{KL},
\end{equation}
where $\alpha, \beta > 0$ are regularizing hyperparameters.

\section{Method}

In this project, we build a QCBNN that is capable to process a real-world medical datasets, such as ultrasound images known as BreastMNIST~\cite{medmnistv2}, as well as deliver reliable uncertainty estimations. In the following text, we outline the details of our implementation and specific performance metrics that are used to evaluate the models.

\subsection{Use-case}
We focus on BreastMNIST~\cite{medmnistv2} dataset, which consists of breast ultrasound $28 \times 28$ images that are classified into malignant and non-malignant categories. This dataset is interesting because it captures the complexity of identifying cancerous samples in a scarce unbalanced setting, as it consists of a mere 780 samples for training, validation and testing. The data is normalized as the only further preprocessing step.

\subsection{Architecture} \label{sec:architecture}

\begin{figure}
    \centering
    \includegraphics[width=0.4\textwidth]{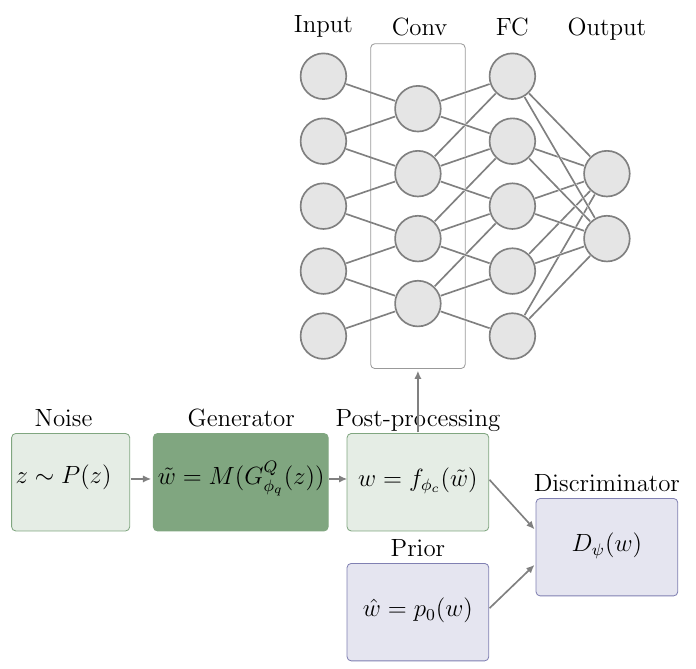}
    \caption{General architecture of proposed continuous hybrid quantum-classical BNN model. The top gray part represents a classical CNN architecture, while the bottom green part represent the quantum weights sampler and the bottom blue part represents the trainers for the quantum sampler. The quantum sampler comprises multiple modules: classical noise, a PQC generator and a post-processing step that computes expectation values of each qubit. The only part that requires a quantum hardware is the quantum generator, while all the other components are performed classically. The trainers of the sampler include a discriminator and a prior.}
    \label{fig:general_architecture}
\end{figure}

We start with an architectural idea that proved useful in a hybrid quantum-classical setup on this dataset, namely QCCNN~\cite{matic2022quantumclassical} (see \cref{sec:related}). We create a Quantum-Classical (Convolutional) BNN (QCBNN) by changing the function of a PQC from performing a convolutional operation to sampling stochastic weights for the classically performed convolution. A sketch of a full architecture is presented in \cref{fig:general_architecture}. The training strategy is adapted from binary quantum BNN as described in \cref{sec:binary_BNN}. The model consists of a quantum \textit{generator} for sampling weights and trainer modules, such as a \textit{prior} and a \textit{discriminator}, that attempts to distinguish quantumly generated samples from the prior. To boost expressivity of the model we adapt a technique for sampling from continuous weights described in \cref{sec:VI_on_QC}, which introduces a classical \textit{noise} and a \textit{postprocessing} layer to our architecture. 

\begin{figure*}
    \centering
     \begin{subfigure}[b]{0.4\textwidth}
        \includegraphics[height=2.3cm]{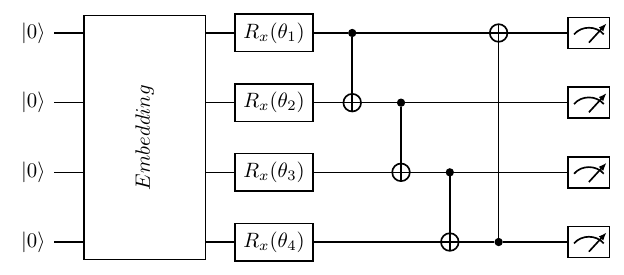}
        \caption{\textit{Matic I} entangling layer \cite{matic2022quantumclassical}}
        \label{fig:matic_i}
    \end{subfigure}~
    \begin{subfigure}[b]{0.4\textwidth}
        \includegraphics[height=2.3cm]{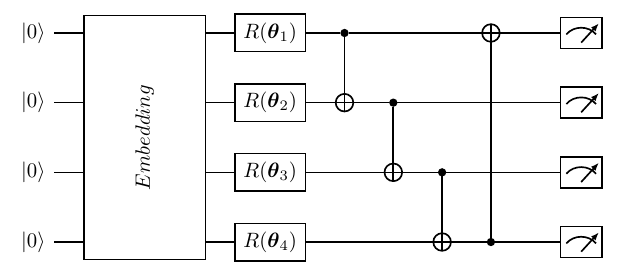}
        \caption{\textit{Matic II} strongly entangling layer \cite{matic2022quantumclassical}}
        \label{fig:matic_ii}
    \end{subfigure}
    \begin{subfigure}[b]{0.4\textwidth}
        \includegraphics[height=2.3cm]{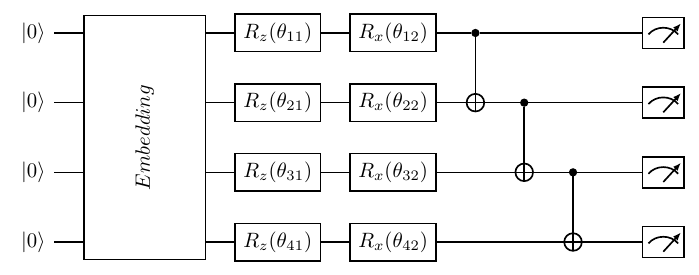}
        \caption{\textit{Nikoloska} inspired by\cite{binary_qBNN}}
        \label{fig:nikoloska}
    \end{subfigure}~
    \begin{subfigure}[b]{0.4\textwidth}
        \includegraphics[height=2.3cm]{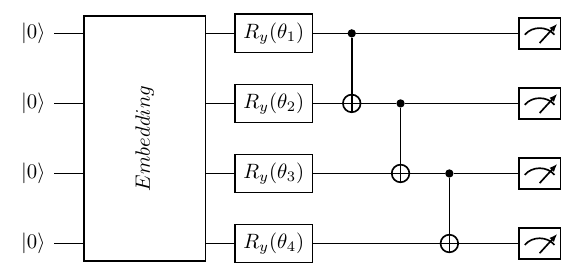}
        \caption{\textit{Romero} \cite{cont_distribution}}
        \label{fig:romero}
    \end{subfigure}
    \begin{subfigure}[b]{0.4\textwidth}
        \includegraphics[height=2.3cm]{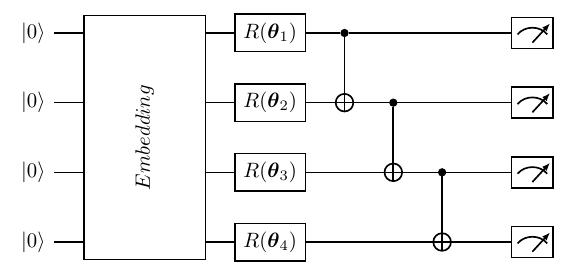}
        \caption{\textit{Circuit I}}
        \label{fig:circuit_i}
    \end{subfigure}~
    \begin{subfigure}[b]{0.4\textwidth}
        \includegraphics[height=2.3cm]{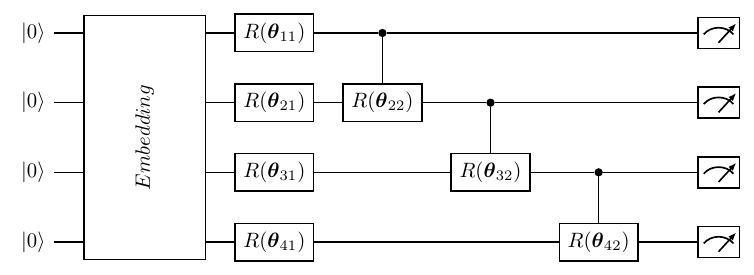}
        \caption{\textit{Circuit II}}
        \label{fig:circuit_ii}
    \end{subfigure}
    \begin{subfigure}[b]{0.4\textwidth}
        \includegraphics[height=2.3cm]{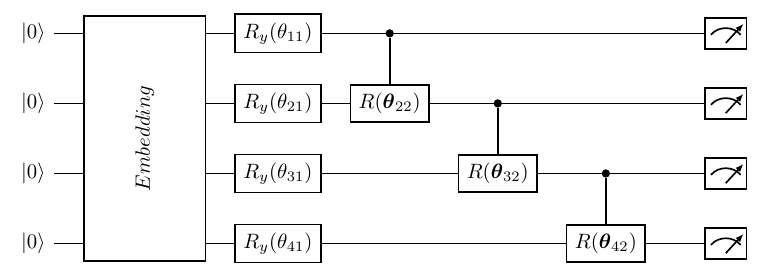}
        \caption{\textit{Circuit III}}
        \label{fig:circuit_iii}
    \end{subfigure}~
    \begin{subfigure}[b]{0.4\textwidth}
        \includegraphics[height=2.3cm]{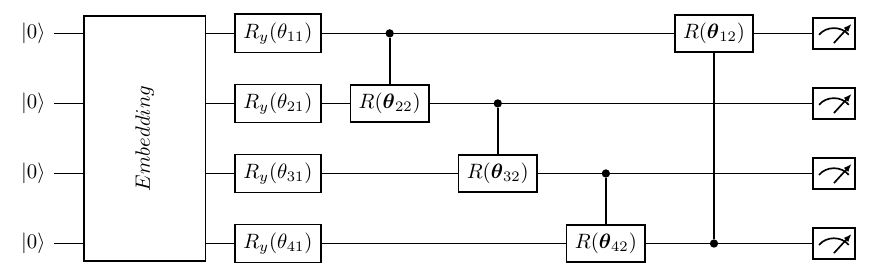}
        \caption{\textit{Circuit IV}}
        \label{fig:circuit_iv}
    \end{subfigure}
    \caption{Different PQC architectures, inspired by the literature. All architectures contain an embedding block, trainable block with rotation gate $R$ parametrized by $\theta$ followed by measurement of all qubits.}
    \label{fig:calculation_layers}
\end{figure*}

The PQC structure in this setup requires two architectural decisions: an \textit{embedding layer} that encodes a classical noise vector and a \textit{calculation layer}. Here, we select an embedding strategy that showed the most promise in other works \cite{Abbas_2021, matic2022quantumclassical} and our internal experiments, namely a \textit{higher-order embedding} layer. The structures of calculation layer, however, are still heavily debated in the literature. Hence, we test out multiple variations of these layers to see which architectural features of PQCs have the most impact on the performance of the overall model. We first test models that were presented in the literature, such as \textit{Matic I}~(\cref{fig:matic_i}) and \textit{Matic II}~(\cref{fig:matic_ii}) (the best and the second best architectures from \cite{matic2022quantumclassical}), \textit{Nikoloska}~(\cref{fig:nikoloska}) inspired by the PQC from binary QBNN work \cite{binary_qBNN} and \textit{Romero}~(\cref{fig:romero}) from the paper on sampling from continuous distributions \cite{cont_distribution}. Additionally, we implement tailored architectures \cref{fig:circuit_i}-\ref{fig:circuit_iv}, which are explained in more detail below.

The classical part is implemented with \texttt{PyTorch} and consists of a one convolutional layer and one fully-connected layer. The weights of the convolutional layer (16 feature maps produced by $2\times2$ kernels) are sampled from a distribution either produced by a classical NN (benchmark) or a PQC. PQC consist of a 4-qubits are implemented in \texttt{PennyLane}. We use \texttt{lightning.qubit} noiseless simulator to simulate PQCs. 

\subsection{Classical benchmark}
For the classical benchmark we implement the same adversarial training loop as for the hybrid model. We substitute the PQC with a minimal width and depth NN that reached comparable performance to conventional VI (see \cref{sec:VI_vanilla}). This NN consists of a single hidden layer of 8 neurons and 4 input neurons (to match 4 qubit setup). 

\subsection{Performance metrics} \label{sec:performance_metrics}

To measure the success of the model, we need to track both, its predictive power as well as the corresponding uncertainty of the predictions. To gauge the predictive performance of the model we track its training and validation accuracy curves and we record additional statistics on the test dataset, such as precision, recall and F1 score\footnote{A harmonic mean of precision and recall, which is a common metric choice for imbalanced datasets}. To assess an uncertainty estimation of the model, we track the mean confidence of the final prediction, which allows us to estimate the model's calibration and confidence error defined as
\begin{equation}\label{eq:confidence}
    \text{Confidence error} = \text{Mean confidence} - \text{Accuracy}.
\end{equation}
We strive to create a model that provides a more reliable information on its certainty in the answer. We therefore compare the confidence of both, correctly and incorrectly estimated samples. Additionally, we track the size of an ensemble that contributed towards the final prediction.

\section{Results}
In this section, we analyse the performance of QCBNN against its classical counterpart. The effects of different PQC architectures on the predictive performance and the uncertainty awareness are tested. We perform this analysis in an iterative fashion as we start (\cref{sec:literature_all}) with the investigation of different architectures from the literature (\cref{fig:matic_i}-\ref{fig:romero}) and bring forward hypotheses on which architectural elements might contribute to their successes and failures. We test these hypotheses further (\cref{sec:custom_all}) as we create custom architectures (\cref{fig:circuit_i}-\ref{fig:circuit_iv}) with emphasised properties to see whether these modifications can elevate the performance. 

\begin{figure*}[h]
\centering
     \begin{subfigure}[b]{\textwidth}
         \centering
         \includegraphics[width=\textwidth]{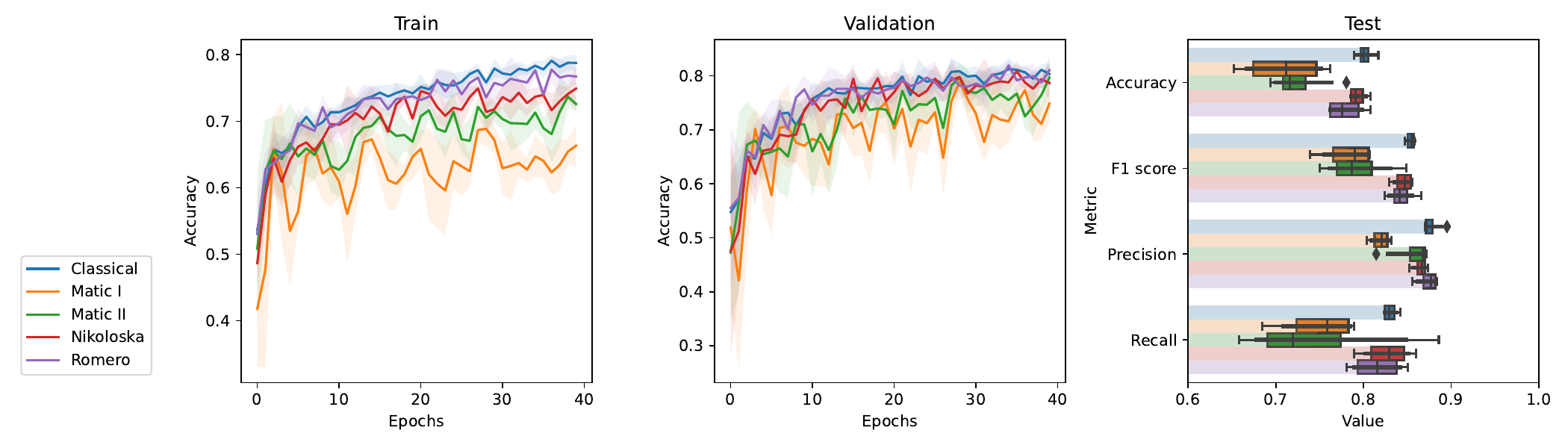}
         \caption{Training (left), validation (center) curves and final test performance (right). Lines of the curves (left and center) indicate mean performance and transparent sleeves indicate standard deviation over four random seeds. The box plots of the final metrics (right) show the quartiles of performances generated by the seeds, while whiskers indicate the spread of the distribution.}
         \label{fig:ZZfeature_performance}
     \end{subfigure}
     \begin{subfigure}[b]{\textwidth}
         \centering
         \includegraphics[width=0.97\textwidth]{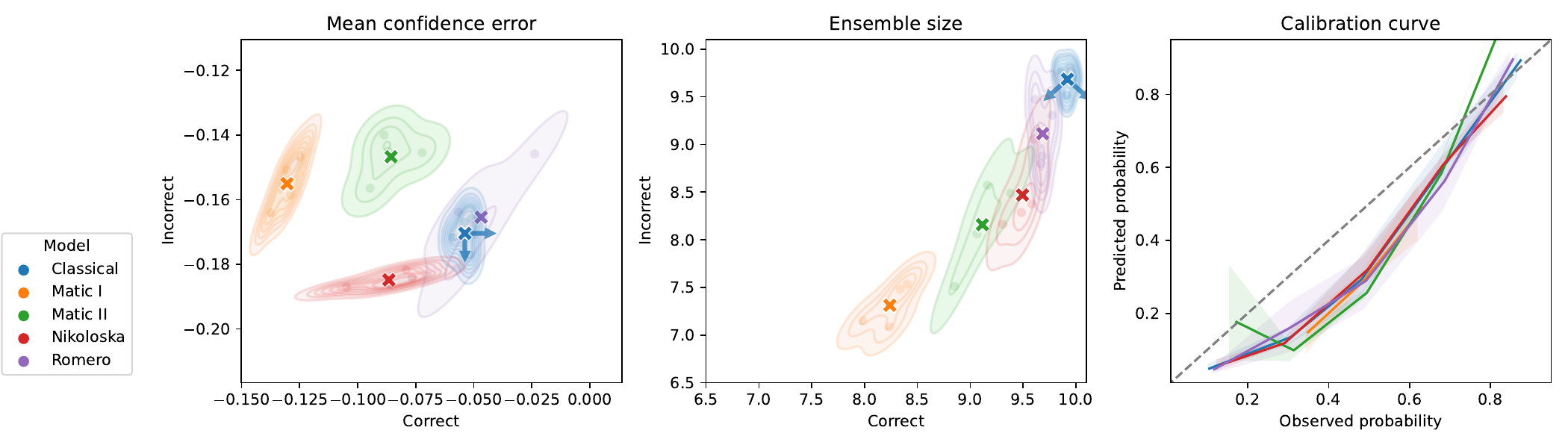}
         \caption{Mean confidence error (left), ensemble size (center) and calibration (right) on test set. Kernel density estimate plots (left and center) show estimated probability distribution of final uncertainty estimation metrics over four random seeds. The X-markers show the means of the distributions (blue marker shows mean of the classical benchmark), while the arrows show the direction of the desired development.}
         \label{fig:ZZfeature_uncertainty}
     \end{subfigure}
    \label{fig:ZZfeature_all}
    \caption{Overview over predictive performance metrics and uncertainty-related metrics for \cref{fig:matic_i}-\ref{fig:romero} PQCs.}
\end{figure*}

\subsection{Architectures from the literature}\label{sec:literature_all}
We first investigate how well a simple setup with a high-order embedding and a varying single calculation layer, that was presented in the literature for similar setups (see \cref{fig:matic_i}-\ref{fig:romero}), performs in the QCBNN on the BreastMNIST dataset. We first consider different performance metrics that capture predictive power and uncertainty awareness separately, then we propose a way to fuse these metrics that enables easier analysis. 

\subsubsection{Predictive performance}
\cref{fig:ZZfeature_performance} shows training and validation curves and final scores (accuracy, F1 score etc.) of \cref{fig:matic_i}-\ref{fig:romero} PQC architectures as well as the performance of the classical benchmark in this setup. All tested architectures struggled to reach the performance of classical benchmark. On the test set all PQCs show more volatile behaviour on recall (model's ability to spot all cancerous datapoints), which seem to be more sensitive to starting conditions than the benchmark.
Interestingly, the PQC architectures \textit{Matic I} and \textit{Matic II} that were used successfully in an analogous setup \cite{matic2022quantumclassical} but performed a different task struggled to learn here. This is visible on both, the highly volatile training and validation curves as well as low final performance. The architectural difference between these PQCs are minor and consist of a mere constraint of rotation, meaning that \textit{Matic~I} only has $R_x$ gates while \textit{Matic~II} allows to fully parametrized the rotation. Changing the entangling strategy by dropping the last entangling gate, as in \textit{Nikoloska} and \textit{Romero}, creates a behaviour that is closer to the classical benchmark on this dataset. Yet, the training and validation curves are still more volatile and stay below the benchmark. Adding one more layer of (phase) rotation gates, as in \textit{Nikoloska}, seems to make the training curves a little more volatile, however, improves test mean accuracy and recall scores.

\subsubsection{Uncertainty estimation}
\cref{fig:ZZfeature_uncertainty} exposes a different side of the performance metrics that are relevant for BNNs, namely uncertainty metrics. As can be seen from calibration curve in \cref{fig:ZZfeature_uncertainty} (right), all models tend to be underconfident for a specific interval (20-80\%) of probabilities, while the \textit{Matic~II} architecture gravitates towards overconfidence on lower and higher probabilities. The mean confidence error as defined in \cref{eq:confidence} (left) and ensemble size plots (middle) show approximate distribution of observations computed by kernel density estimate. The means of these distributions are indicated by crosses and blue errors indicate the direction of a favourable development in comparison to a classical benchmark. We track these values for both correctly and incorrectly identified samples to visualize how confidently wrong the models are. For the mean confidence error, we would like to see a lower (absolute) confidence error for correct samples and a higher confidence error on incorrect samples, which would indicate a higher uncertainty. All models stay in the negative range that indicate underconfidence of all models on the final prediction, which is consistent with the calibration curve readings (right plot). For the ensemble size, we would like to have a higher number of ensemble members to vote for a prediction that was correct than incorrect. Interestingly, classical learners' ensemble members seem to converge to a homogeneous behaviour and vote the same way no matter what. For quantum learners we see more variety. \textit{Matic I} shows a similarly high confidence error for both correctly and incorrectly identified samples. While \textit{Matic II} improves the mean confidence error, it still differentiates itself substantially from the classical benchmark. Both of these model also show a lesser ensemble size for both correct and incorrect samples. Together, these metrics might hint towards poor learning during the training phase, which was already discussed above. \textit{Nikoloska} showed higher uncertainty for incorrect samples while still exhibiting overall high underconfidence. \textit{Romero}'s mean confidence error distribution, on the other hand, overlapped with the one from the classical benchmark, however had a bigger spread.

\subsubsection{Weights distribution}
\begin{figure}[h]
    \centering
    \includegraphics[width=0.35\textwidth]{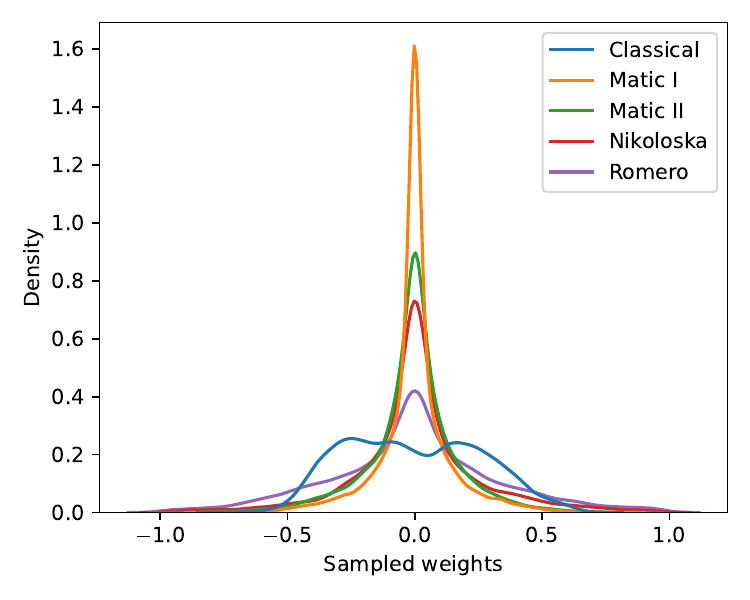}
    \caption{Weights distribution density estimated from 100 inference passes through the model and consists of all samples across all convolutional filters combined.}
\label{fig:weights_entanglement}
\end{figure}

Now, we take a closer look at how the output of the PQC, which generates the weights distribution, changes as we modify the architectures. \cref{fig:weights_entanglement} shows that all of PQC's expectation values are biased towards zero. The architectures that had more wide spread distributions tend to have stronger performances in terms of metrics considered above. Keeping this in mind, we proceed to tweaking the architectures in the following sections to see if we can force the weights distributions to take a more expressive form and see whether this maps well to improved performance.

\subsubsection{Choosing architecture}\label{sec:literature_difference}

\begin{figure}
         \centering
    \includegraphics[width=0.35\textwidth]{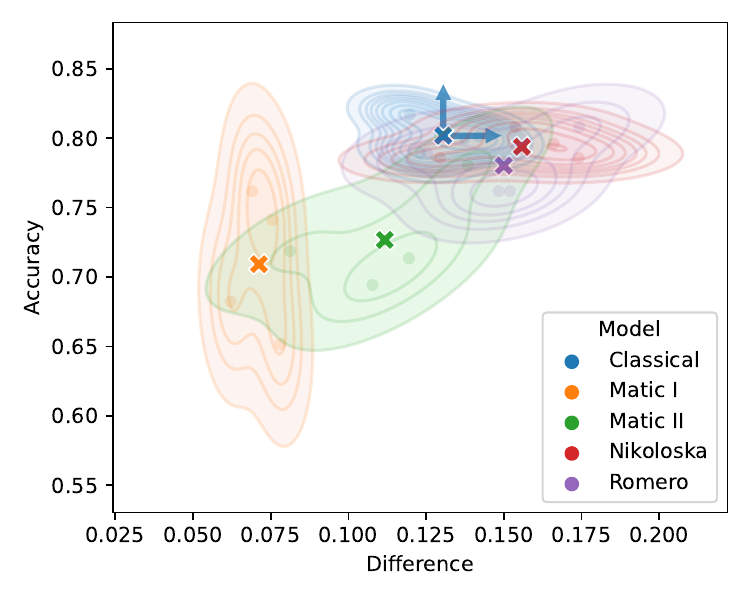}
    \caption{Accuracy on the test set and certainty level difference between correctly and incorrectly identified samples weighted by ensemble size (see \cref{eq:difference}).}
    \label{fig:literature_difference}
    \end{figure}

To find the best performing PQC we need to consider multiple behavioural parameters at once. To ease the analysis, we fuse the uncertainty metrics and compare it to the final accuracy. Given that quantum models show a bigger discrepancy between the ensemble size of correct and incorrect predictions, we weight the mean uncertainty with the ensemble size to acquire a more reliable uncertainty estimation. It is our ambition to find models that posses more dependable point-wise uncertainty estimation capabilities without sacrificing predictive accuracy. We therefore consider whether a PQC provides for a bigger difference in confidence between correctly and incorrectly identified samples. This difference is computed as follows:
\begin{align} \label{eq:difference}
    \text{Difference} & = \text{Confidence}_{\text{correct}} \times \text{Ensemble fraction}_{\text{correct}} \nonumber\\
    & - \text{Confidence}_{\text{incorrect}} \times \text{Ensemble fraction}_{\text{incorrect}}
\end{align}
\cref{fig:literature_difference} shows the estimated distribution of these differences versus accuracy of prediction for all PQC architectures considered above. This graph indicates that the PQCs that showed the best performances above (\textit{Nikoloska} and \textit{Romero}) are showing on average a slightly higher confidence difference between correctly and incorrectly identified cases at the expense of a drop in accuracy compared to the benchmark. The distributions of these architectures are significantly overlapping, but \textit{Nikoloska} seems to have a more stable performance in terms of accuracy and slightly higher average difference. This architecture, however, also showed a higher mean confidence error (meaning that the model was less confident in its predictions on average) as can be seen \cref{fig:ZZfeature_uncertainty}, which is not reflected in the difference metric.  

From the above analysis we conclude that \textit{Romero} architecture had one of the strongest predictive performances on all training, validation and test datasets, the lowest average mean confidence error for the correct prediction, one of the largest confidence gap between correct and incorrect samples, and the most wide spread of the weights distribution generated.

\begin{figure*}[h]
 \centering
     \begin{subfigure}[b]{\textwidth}
         \centering
         \includegraphics[width=\textwidth]{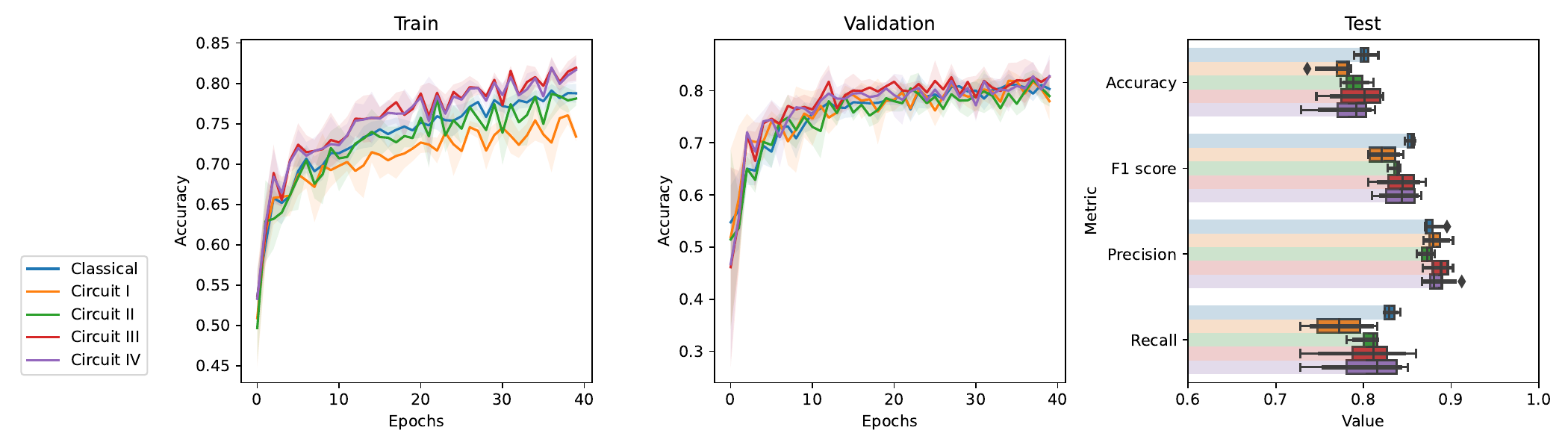}
         \caption{Training (left), validation (center) curves and final test performance (right). Lines of the curves (left and center) indicate mean performance and transparent sleeves indicate standard deviation over four random seeds. The box plots of the final metrics (right) show the quartiles of performances generated by the seeds, while whiskers indicate the spread of the distribution.}
         \label{fig:custom_performance}
     \end{subfigure}
     \begin{subfigure}[b]{\textwidth}
         \centering
         \includegraphics[width=0.97\textwidth]{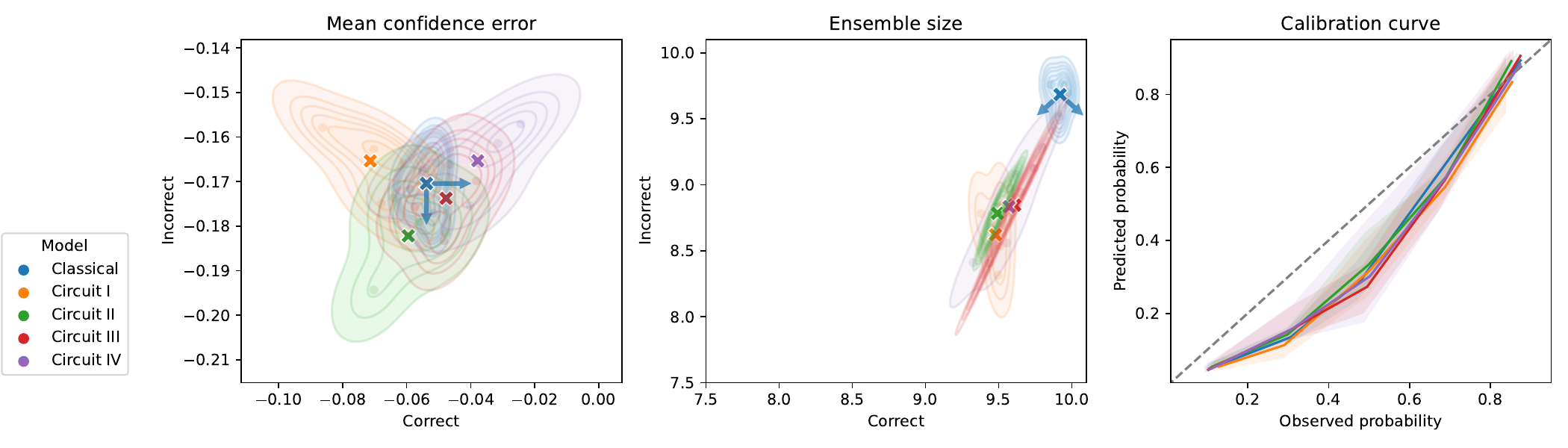}
         \caption{Mean confidence error (left), ensemble size (center) and calibration (right) on test set. Kernel density estimate plots (left and center) show estimated probability distribution of final uncertainty estimation metrics over four random seeds. The X-markers show the means of the distributions (blue marker shows mean of the classical benchmark), while the arrows show the direction of the desired development.}
         \label{fig:custom_uncertainty}
     \end{subfigure}
    \label{fig:custom_all}
    \caption{Overview over predictive performance metrics and uncertainty-related metrics for \cref{fig:circuit_i}-\ref{fig:circuit_iv} PQCs.}
\end{figure*}

\subsection{Custom architectures}\label{sec:custom_all}

PQC architectures that were previously proposed in the literature, unfortunately, failed to deliver favourable results in comparison to classical benchmark for the task at hand as was discussed above. Moreover, the distribution of weights that quantum samplers generated had a strong bias towards $0.$, which challenges the common argument for using QC for probabilistic ML such as representing distributions that are hard to model classically. Here, we built up upon the insights from the previous section in an attempt to create stronger architectures for our problem. Allowing the architectures a full instead of partial control, as in going from \textit{Matic~I} to \textit{Matic~II}, boosted the predictive capability. Therefore, we perform the same trick with the best performing architecture from the literature \textit{Romero} and create \textit{Circuit I} (\cref{fig:circuit_i}). Another acquired insight from the previous study is that the change in the entanglement layer had the most profound impact on the PQC's behaviour (moving from \textit{Matic I} and \textit{Matic II} to \textit{Nikoloska} and \textit{Romero}). It was beneficial for both, predictive performance and uncertainty awareness. Therefore, we change the entangling layer from CNOT to CR gates for \textit{Romero} and \textit{Circuit I} and create \textit{Circuit III} (\cref{fig:circuit_iii}) and \textit{Cirucit II} (\cref{fig:circuit_ii}) respectively to allow the model to tune its own entangling during training. Additionally, we include \textit{Circuit IV} where we add an entangling gate between first and last qubit to test whether this entanglement strategy is generally not fitting for this task or not. 

\subsubsection{Predictive performance}
\cref{fig:custom_performance} shows the performance of different custom architectures (\cref{fig:circuit_i} - \cref{fig:circuit_iv}). Unexpectedly, providing more freedom to the PQC to tweak its rotation with \textit{Circuit I} hindered the training in comparison to \textit{Romero} (see \cref{fig:ZZfeature_performance}), which seems to indicate that there is a possible saturation point beyond which a free choice of rotation is no longer boosting the model's capacity. Proving more freedom for the entangling layers, on the other hand, lead to the expected boost: \textit{Circuit II} provided an improved performance over \textit{Circuit I} and previously tested PQC in \cref{fig:ZZfeature_performance}. Its performance on training, validation and test set is, however, still on average a bit lower than of the classical benchmark. Restricting the rotation layer the same way as in \textit{Romero} (\cref{fig:romero}) for architectures \textit{Circuit~III} and \textit{Circuit~IV} provided enough capacity to go slightly beyond the classical benchmark on the training dataset. Here, we see barely any difference between two entangling strategies in \textit{Circuit~III} and \textit{Circuit~IV}. Interestingly, even though there is a clear distinction of the training curves on the training data, the ones on validation data show almost no discrepancy. On the test dataset, \textit{Circuit~III} managed to slightly outperform the classical benchmark in terms of the average accuracy, but the spread of accuracies is high. Looking at other metrics, we can see that quantum models tend to provide a higher precision (high quality of solution), but suffer from wide spread in recall, meaning that, dependent on initial conditions, our model managed to either over- or underperform in comparison to the classical benchmark. 

\subsubsection{Uncertainty estimation}
\cref{fig:custom_uncertainty} provides an overview of different uncertainty measures of the model. Here again, as in \cref{fig:ZZfeature_uncertainty}, we see barely any difference in calibration curves of the models. The mean confidence error distributions, however, tell a slightly different story: \textit{Circuit I} that had trainability issues, as discussed above, also shows lower mean confidence error on correct samples and higher error on incorrect samples. \textit{Circuit II} on average shows a slightly higher uncertainty on incorrectly identified samples, without rising the uncertainty of the correctly identified samples. This analysis is valid on average, but given the spread of the error distribution, each single model might show a bit different behaviour. Constraining the rotation layer of the previously discussed model to \textit{Circuit~III} leads to lesser spread of the error distribution and lesser mean confidence error for both correctly and incorrectly identified samples. Changing the entanglement strategy from \textit{Circuit III} to \textit{Circuit IV} leads to a smaller on average mean confidence error for both correctly and incorrectly classified instances. In terms of ensemble size we see a favourable development for all quantum models as the correct samples receive more votes in favour, while for incorrect samples ensemble agree less.

\subsubsection{Weights distribution}
\begin{figure}[h]
     \centering
\includegraphics[width=0.35\textwidth]{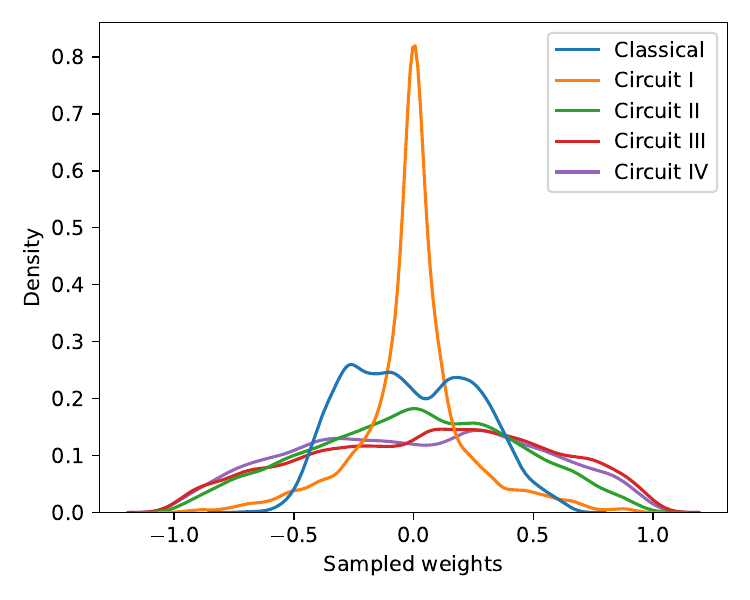}
\caption{Weights distribution density estimated from 100 inference passes through the model and consists of all samples across all convolutional filters combined.}
\label{fig:custom_weights}
 \end{figure}

Now, we analyse the distributions that were generated by the PQCs. \cref{fig:custom_weights} provides an overview of the effects of different architectures on the actual output of the PQCs. \textit{Circuit I}'s output that had the most constrained entanglement regime is still biased towards $0.$ (the same as for architectures analysed previously), however, changing the entanglement layer allowed the distributions to be wider and take more expressive forms. From this graph we see that the models that performed at least the same as a classical benchmark also have at maximum the same density level, which aligns with a hypothesis from above that BNN favour generators that provide diverse outputs.

\subsubsection{Choosing architecture}
\begin{figure}[h]
    \centering
    \includegraphics[width=0.35\textwidth]{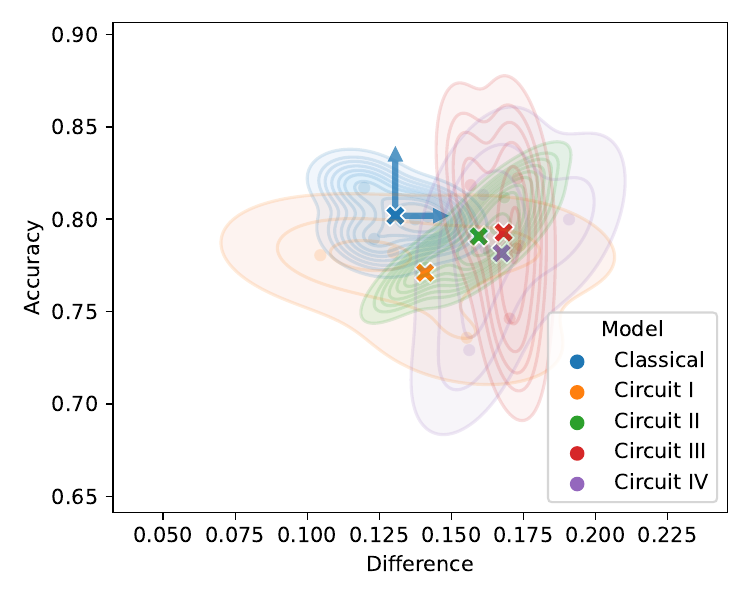}
    \caption{Accuracy on the test set and certainty level difference between correctly and incorrectly identified samples, weighted by ensemble size (see \cref{eq:difference}).}
    \label{fig:custom_difference}
\end{figure}

Here, we perform the same analysis as was done in \cref{sec:literature_difference} for tweaked PQC architectures \cref{fig:circuit_i} - \ref{fig:circuit_iv}. \cref{fig:custom_difference} shows the distributions of differences and accuracies across different architectures. All quantum architectures show on average a slightly higher difference between correctly and incorrectly identified samples, which comes at an expense of a lower average accuracy. The lowest performing PQC \textit{Circuit I} shows a wide spread of the difference distribution. Relaxing the entanglement layers in \textit{Circuit II} leads to a much more narrow distribution, which reaches the differences at least as good as the classical benchmark. The average predictive performance is, however, slightly lower. Changing the entanglement strategy as in \textit{Circuit~III} leads to a higher difference, which overlaps with classical benchmark only a little. The predictive performance of this model greatly depends on initial conditions as discussed above. Adding one additional entanglement gates as in \textit{Circuit~IV} slightly lowers the average performance and increases the spread of the distribution.

Based on results from above, we draw the conclusion that PQC architecture \textit{Circuit III} had the strongest performance compared to \textit{Circuit I, II} and \textit{IV}, as well as PQCs \textit{Matic I, Matic II, Nikoloska} and \textit{Romero} from the previous section. \textit{Circuit III} also slightly exceeded the classical benchmark in predictive performance and uncertainty estimation. To test whether the performance of \textit{Circuit III} has saturated we performed additional testing in \cref{sec:layers_all}.

\section{Discussion}
The results described in the previous section imply that by introducing learnable parameters into the entangling layer allows to achieve better performances for this task, which is not a common practice that we see across the literature. It correlates with weights distributions having a wider spread, which seem to be more favoured by the setup of a QCBNN. On the contrary, in the setup of QCCNN~\cite{matic2022quantumclassical} the PQCs \textit{Matic I} outperformed \textit{Matic~II}, while in our setup these PQCs are the worst performing ones and fall below the classical benchmark. This could indicate that the tasks of computing convolutional operation (as in QCCNN) and sampling from weights distribution (as in QCBNN) favour contrasting behavioural patterns of the models.

An interesting further research direction is to provide more theoretical basis that would allow to construct tailored PQCs in a less heuristic fashion. One way to approach this is to investigate the connection of these circuits to the coefficients of Fourier series that can be fitted to the periodic output of the PQCs as proposed in \cite{Schuld_2021}. In that work, the authors showed that the embedding of a circuit controls the number of frequencies in the series and thus the expressivity of the model (the more frequencies are available, the more complex function one can represent with the models). However, it is speculated that the structure of the computation layer effects the coefficients of these series, but this idea was not developed further. In our work, we have shown that some architectural decisions of PQCs resulted in more wide distributions of the weights, and it might prove to be beneficial to investigate this point more thoroughly by looking into the connection of these decisions to the coefficients. A related further research direction is a connection of the PQCs to classical surrogates as introduced in \cite{Schreiber_2023}. Provided that the structure of selected PQCs fulfills the criteria, it is possible to build a classical surrogate model that could serve as a more reliable classical benchmark as well as ease deployability of these models. Allowing too much freedom in rotational parameters is, however, detrimental to the performance (e.g. \textit{Circuit II} to \textit{Circuit III}). This could be attributed to more complex loss landscapes that hinder the optimizer to find optima. However, a deeper dive into the origins of the problem might shed some light onto why some PQC architectures are more successful than others.

One things that becomes apparent through out experiments described above is that the training curves of the PQC in \cref{fig:ZZfeature_performance} and \cref{fig:custom_performance} do not necessarily map well to the accuracies of these architectures on the test dataset, which could be attributed to a rather small test dataset. In future work, we are planning to perform further experiments to test how these results map to bigger datasets. Furthermore, in this work, we do not distinguish between epistemic and aleatoric uncertainties, like it was done in e.g. \cite{9759399}, which is an interesting additional metric to consider in the future works.

In a pursuit of allowing QCBNN model to represent continuous weights and hence learn more complex datasets we modified Born machine according to \cite{cont_distribution}. This entailed that the source of stochasticity was shifted from quantum to classical devices. A further interesting direction of research could be to change the way we sample from the generator to allow the PQC to remain a Born machine and yet keep the possibility for weight to remain continuous. One idea to do this would be to implement something akin Monte Carlo Dropout methods, where PQC can be responsible for selecting the weights that are dropped out during this iteration. There are a few challenges to overcome with this idea, e,g. dropout is often implemented as a Bernoulli distribution, so what type of PQC architecture could be beneficial here and whether or not it should have learnable parameters remains to be investigated.

\section{Conclusion}

In this work, we introduce a Quantum-Classical Bayesian Neural Network (QCBNN) with continuous weights for uncertainty-aware classification of breast ultrasound scans (BreastMNIST). This model is designed as a fusion of a previously introduced models \cite{matic2022quantumclassical,binary_qBNN, cont_distribution} and it is evaluated from the stand points of its predictive performance and uncertainty awareness. Many behavioral aspects of the models are considered that are captured by a variety of metrics. We methodically test multiple PQC architectures in this scenario, including the ones that has been successfully used in a similar setup before (\cref{fig:matic_i}-\ref{fig:romero}). We determined that certain architectural features, such as trainable entanglement layers and rotation layers with less parameters, allow for better learning on this dataset. We test these hypotheses by building custom PQCs for this task (\cref{fig:circuit_i}-\ref{fig:circuit_iv}) that boost the performance and even allow to slightly outperform the classical benchmark in terms of uncertainty awareness. These results provide interesting empirical insights for more informed PQC designs in the future.

\section{Acknowlegements}

This research is supported by the Bavarian Ministry of Economic Affairs, Regional Development and Energy with funds from the Hightech Agenda Bayern. We thank Marc Machaczek for his role in the preparatory stage of the project.


\printbibliography
\begin{appendices}

\section{Different number of layers for Circuit III}\label{sec:layers_all}

Here, we investigate whether the performance of best performing \textit{Circuit~III} presented in \cref{sec:custom_all} can benefit from deepening its architecture. The layers can be added by simple repeating the computation layer, or adding both embedding and computation layers together to create a re-uploading model~\cite{P_rez_Salinas_2020}. 

\begin{figure}[h]
     \centering
    \includegraphics[trim={0 0 22cm 0.8cm},clip, width=0.42\textwidth]{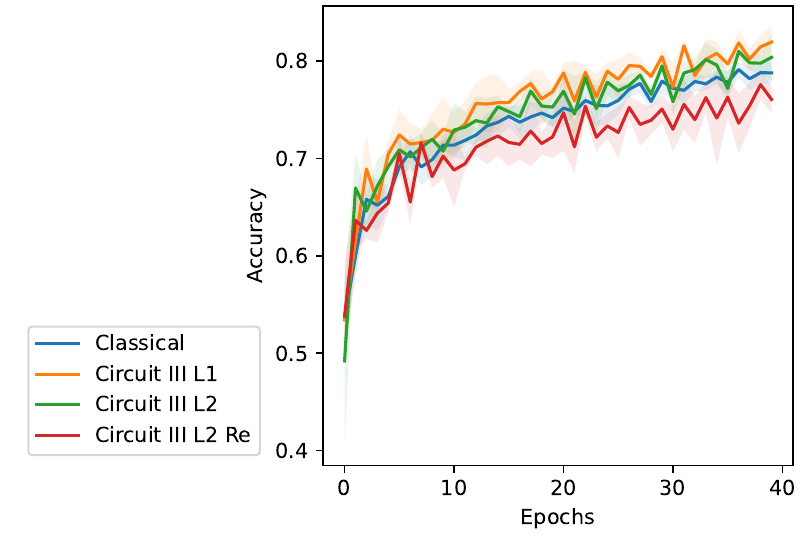}
    \caption{Training curves for \cref{fig:circuit_iii} PQCs with varying depth. Lines of training and validation curve indicate mean performance and transparent sleeves indicate standard distribution over four random seeds.}
    \label{fig:layers_III}
\end{figure}

\cref{fig:layers_III} shows predictive performance of  \textit{Circuit III}~(\cref{fig:circuit_iii}) with one layer (\textit{L1}, the same setup as in \cref{sec:custom_all}), with two layers (\textit{L2}) and two layers in re-uploading setup (\textit{L2 Re}). Increasing the depth from \textit{L1} to \textit{L2} seems to affect the trainability as the average training curve falls below one layer setup. Adding additional embedding layer before \textit{L2} (\textit{L2 Re}) seems to hinder the trainability even further, as performances on training, validation and test datasets are impaired. 

\begin{figure}[h]
    \centering
    \includegraphics[width=0.35\textwidth]{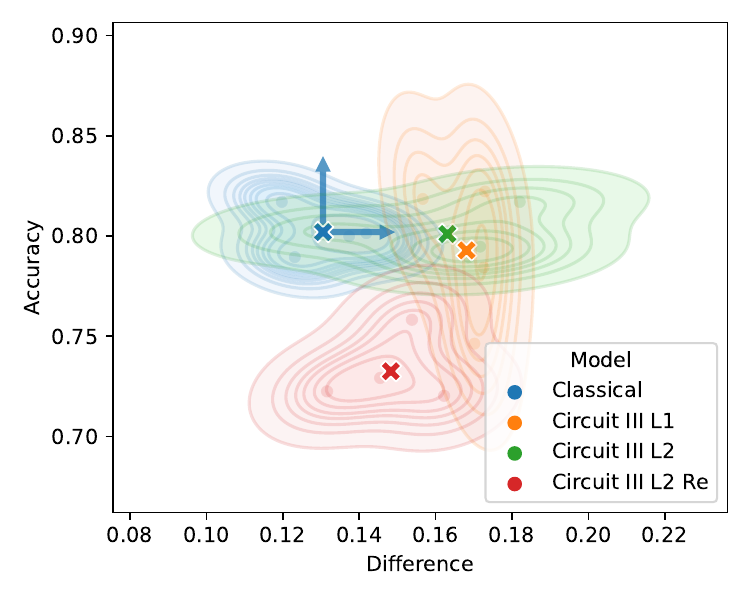}
    \caption{Prediction performance as measured by accuracy on the test set and uncertainty estimation difference between correctly and incorrectly identified samples that is weighted by ensemble size (see \cref{eq:difference}).}
    \label{fig:layers_difference}
\end{figure}

Next, we consider test accuracy and difference confidence levels between correctly and incorrectly identified samples (as in \cref{eq:difference}). \cref{fig:layers_difference} displays these metrics for different number of layers. One layer \textit{Circuit III L1} PQC shows higher confidence gap but quite volatile predictive performance as final accuracy between four random seeds differ a lot. Adding another computation layer (\textit{L2}) stabilizes the predictive performance to the levels similar to the classical benchmark, but destabilizes the uncertainty estimation capacity. 

From the above analysis we conclude that increasing the amount of layers two \textit{L2} stabilizes the predictive performance on testing set while proving a larger confidence gap on average. The average performance on training dataset on the other hand dropped a little, which could hint towards drop in trainability.

\section{Variational Inference}\label{sec:VI_vanilla}
\begin{figure}[h]
    \centering
    \includegraphics[trim={0 0 22cm 0.8cm},clip, width=0.42\textwidth]{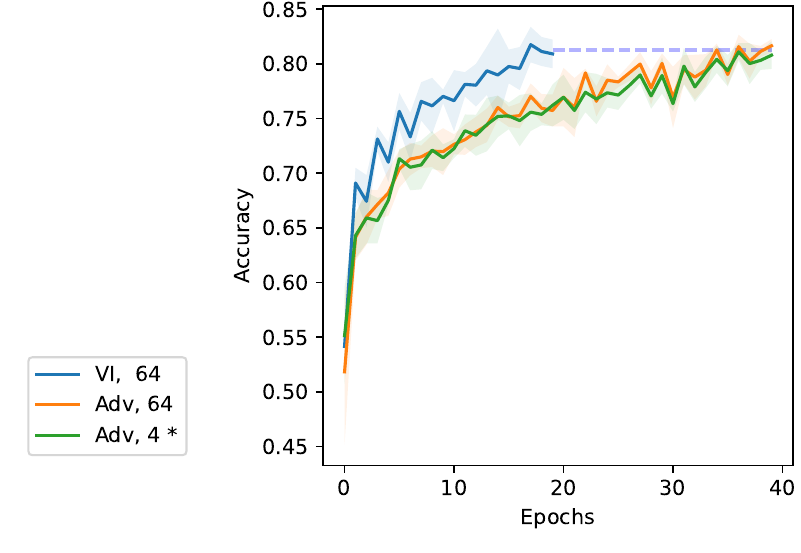}
    \caption{Variational Inference (VI) vs Adversarial learning loop (Adv) training curves of classical BNNs. The numbers next to method name indicate the size of the injected noise vector.}
    \label{fig:VI_vs_Adv}
\end{figure}

Here, we show the comparison between a standard approach to training classical BNNs namely the Variational Inference (VI) and the Adversarial training (Adv) that serve as a classical benchmark. \cref{fig:VI_vs_Adv} shows that Adv achieves the same accuracy as VI method even with substantially smaller injected noise vector, however, it take more epochs to do so.

\end{appendices}

\end{document}